%% file: main.tex
\documentclass[conference]{IEEEtran} 

\input{packages.tex}

\input{macros.tex}

\begin{document}

\title{Using the TypeScript compiler to fix\\erroneous Node.js snippets}

\author{\IEEEauthorblockN{Brittany Reid}
\IEEEauthorblockA{
The University of Adelaide, Australia\\
brittany.reid@adelaide.edu.au
}
\and
\IEEEauthorblockN{Christoph Treude}
\IEEEauthorblockA{The University of Melbourne, Australia\\
christoph.treude@unimelb.edu.au}
\and
\IEEEauthorblockN{Markus Wagner}
\IEEEauthorblockA{Monash University, Australia\\
markus.wagner@monash.edu}
}


\maketitle

\begin{abstract}

Most online code snippets do not run. This means that developers looking to reuse code from online sources must manually find and fix errors. We present an approach for automatically evaluating and correcting errors in Node.js code snippets: Node Code Correction (NCC). NCC leverages the ability of the \ts\ compiler to generate errors and inform code corrections through the combination of \ts{}'s built-in codefixes, our own targeted fixes, and deletion of erroneous lines. Compared to existing approaches using linters, our findings suggest that NCC is capable of detecting a larger number of errors per snippet and more error types, and it is more efficient at fixing snippets. We find that 73.7\% of the code snippets in NPM documentation have errors; with the use of NCC's corrections, this number was reduced to 25.1\%. Our evaluation confirms that the use of the \ts\ compiler to inform code corrections is a promising strategy to aid in the reuse of code snippets from online sources.

\end{abstract}

\section{Introduction}

Most code snippets online do not run; existing work has shown that only 15.2\% of Node.js snippets in NPM package documentation are runnable~\cite{chinthanet2021makes}. Because software developers frequently reuse code from online sources~\cite{Baltes2019}, they often need to dedicate time to fixing errors. This introduces challenges when using third-party libraries: examples in documentation are intended to demonstrate usage and non-working snippets can present a barrier to getting started.

Because code snippets are not full, runnable programs with test cases, existing work in automating the detection and fixing of errors has primarily focused on static analysis~\cite{reidnlp2tc, yang2016query, reidNCQ, campos2019mining, 10.1145/3520304.3528772, 10.1145/3520304.3533970}. Code reuse tools such as NLP2TestableCode~\cite{reidnlp2tc} and NCQ~\cite{reidNCQ} combine error detection with heuristic fixes and line deletion to aid developers in reusing snippets. This use of line deletion aims to reduce snippets to an optimal form through a simple deletion operation, looking at errors to determine if the change should be `accepted'. Static analysis is also useful for measuring the \textit{quality} of code; existing code reuse tools have made use of parsers, linters and compilers to report errors and find the `best' snippet for a given search query~\cite{reidnlp2tc, reidNCQ}. Additionally, such tools can provide insights on the quality of online code in general: for example, Yang et al.~\cite{yang2016query} looked at the usability of Stack Overflow snippets via static analysis. 

Research in Java leverages the compiler for error detection and correction~\cite{terragni2016csnippex, reidnlp2tc}, but \js (and thus the Node.js runtime environment), is an interpreted language that lacks such a compiler. Similar work has instead relied on parsers and linters~\cite{yang2016query, reidNCQ, campos2019mining}. For example, NCQ~\cite{reidNCQ}, a command-line REPL (Read-Eval-Print-Loop) programming environment, which automates the process of reusing code snippets from NPM package documentation, uses ESLint~\cite{eslint} to report errors, and increases the number of snippets without errors from 54.8\% to 94.0\%. However, these tools serve a different purpose than the compiler (formatting code or generating ASTs), so the errors reported may be more limited; for example, the majority of ESLint rules are stylistic or best practice, not programming errors that affect runnability~\cite{campos2019mining}. Additionally, both ESLint and the SpiderMonkey parser report only a single error if they fail to parse. ESLint needs to successfully parse a snippet to create an AST and run its rule detection. Unlike a compiler, ESLint does not do any type checking. This reveals the need for a better way to evaluate errors in Node.js code.

We investigate how effective the \ts~\cite{typescript} compiler is for reporting and fixing errors in Node.js snippets, which is a novel contribution to an area that has otherwise relied on linters and parsers. While \ts\ is a superset of \js\ with static typing, the compiler is used in VSCode to provide error highlighting and fix suggestions for \js\ as well~\cite{vscodets}, suggesting it may be more useful for error detection and correction than existing approaches. The existing in-editor implementation requires a degree of manual interaction to handle errors; we take \ts{}'s fix suggestions and apply them automatically on given snippets. Furthermore, we implement a limited set of heuristic fixes targeting the most common errors, leveraging \ts{}'s ability to generate ASTs and provide type information. We present our approach, Node Code Correction (NCC), which adapts NCQ's corrections to use the \ts\ compiler in place of ESLint, including targeted fixes and line deletion. We run both approaches with a dataset of more than two million NPM code snippets and then evaluate NCC against a dataset of Stack Overflow snippet edit pairs representing manual error corrections over time. We report the following findings:

\begin{itemize}[leftmargin=0.5cm, label={*}]
    \item The \ts\ compiler reports more errors than ESLint: on average 6.8 vs 1.3 errors per snippet. ESLint reports a single error and no AST for 47.46\% of erroneous snippets.
    \item \ts\ enables NCC to improve the rate of error-free snippets by 184.67\% compared to 72.60\% for NCQ, with less empty snippets (7.41\% vs 14.33\% of the dataset).
    \item ESLint's built-in fixes had a negligible impact on NCQ's code corrections; only 1 snippet was made error-free. In contrast, \ts{}'s codefixes corrected 79,613 snippets.
    \item 1,099 (6.88\%) of 15,969 Stack Overflow snippets were manually made error free between versions; in comparison, NCC was able to correct 46.77\%. Of this 1,099 that were fixed manually, NCC could fix 66.06\%.
\end{itemize}

\noindent
These results provide evidence that the \ts\ compiler can be useful in automatically identifying and fixing errors, to help reuse online code snippets. We conjecture that further improvement to heuristic fixes can increase the number of corrected snippets. Our approach and related data are available at: \url{https://doi.org/10.5281/zenodo.8272874}

\section{Motivating Example}
\label{sec:example}

A developer wants to read some data from a URL in Node.js. Let us say that the developer comes across the snippet in \autoref{fig:motivating_original} (an unedited Stack Overflow snippet from our dataset) while searching on Google. Like many snippets found online, it has errors.

\begin{figure}[h]
    \centering
    \centering\begin{lstlisting}
http.get(url, function(res) {
    var data = '';
    res.on('data',function(chunk){data+= chunk;});
    res.on('end',function(){
        console.log("BODY: " + data);})
}).on('error', function(e) {
    console.log("Got error: " + e.message);});
};
    \end{lstlisting}
    \caption{Example code snippet from Stack Overflow answer 45582298.}
    \label{fig:motivating_original}
\end{figure}

The developer pastes the snippet into their file, but it fails to run with the error `SyntaxError: Unexpected token '\}'', due to a hanging bracket. Furthermore, the \ts\ compiler identifies a number of other issues with the snippet: the identifier \texttt{url}, and \texttt{http} are also undefined. As we find in \autoref{sec:eval}, these are common errors, as example code often omits parts to simplify the snippet.

\begin{figure}[h]
    \centering
    \begin{lstlisting}[language=diff]
+ const http = require("http");
+ var url = "Your Value Here"; // Suggested Type: 
+        string | RequestOptions | URL
http.get(url, function(res) {
    var data = '';
    res.on('data',function(chunk){data+= chunk;});
    res.on('end',function(){
        console.log("BODY: " + data);})
}).on('error', function(e) {
    console.log("Got error: " + e.message);});
-//};
    \end{lstlisting}
    \caption{Code snippet after NCC's corrections.}
    \label{fig:motivating_ncc}
\end{figure}

However, running NCC before using the snippet results in a snippet that reports no errors, as shown in \autoref{fig:motivating_ncc}. Using \ts{}, NCC detects these errors before without needing to run untrusted code. Using custom fixes, NCC adds the missing \texttt{http} require statement; then for the undefined \texttt{url}, a placeholder value is declared with suggested types to guide the developer. Line deletion then removes the hanging bracket. From this snippet, the developer can make the necessary changes needed to make the snippet runnable.

\begin{figure}[h]
    \centering
    \begin{lstlisting}[language=diff]
+ var https = require('https');
+ var url = 'https://www.alphavantage.co/query...';
+ exports.handler = function (event, context) {
\   https.get(url, function(res) {
    ...});
};
    \end{lstlisting}
    \caption{Excerpt of the manually corrected snippet.}
    \label{fig:motivating_manual}
\end{figure}

We can compare these changes on the original, erroneous version of the snippet, to the manually fixed snippet from our dataset. Similarly, the manually corrected snippet adds the missing \texttt{url} variable, which is a string URL. It also adds a require, but changes the library to \texttt{https} to match the URL. To correct the hanging bracket, the code has been wrapped in an exported function.

In contrast, ESLint reports only a single parsing error for the original snippet, and so the only change is to comment out the bracket. This motivating example illustrates how the capabilities of the \ts\ compiler can be used to help with corrections that benefit developer workflows.

\section{Related Work}

While many automated code reuse tools allow developers to find and use snippets online from within their programming environment, developers still need to spend time correcting them when they do not work. Node Code Correction combines ideas from three areas of work into one tool, in order to help developers reuse code from online: 1) error detection via static analysis; 2) code correction and 3) code deletion. We discuss existing work in these areas, both in the limited \js\ and Node.js space, and in other languages.



\subsection{Static Analysis}

Previous work has looked at error detection in \js\ and Node.js, using parsers~\cite{yang2016query}, linters~\cite{campos2019mining, reidNCQ} or runtime errors~\cite{chinthanet2021makes, yang2016query}. The benefit of static analysis is that it can report multiple errors, is typically fast, and that code can be evaluated without running it; this is especially useful when most online snippets do not run~\cite{chinthanet2021makes, yang2016query}. Additionally, it is undesirable for a code reuse tool to run arbitrary code from online, when snippets can be malicious or contain vulnerabilities.  For these reasons, static analysis can be useful for providing information about large sets of code, or for on-demand use in an automated code reuse pipeline. For example, tools such as NCQ~\cite{reidNCQ} and NLP2TestableCode~\cite{reidnlp2tc} use errors to inform fixes and recommend the highest quality snippets first. However, most of the issues that linters like ESLint report are stylistic; Campos et al. ran the standard ESLint configuration on \js\ code snippets mined from Stack Overflow and found that no snippets were free of rule violations, but that 163 rules could be characterised as `stylistic issues' or `best practice'~\cite{campos2019mining}.

Similar work in Java has made use of static analysis tools like PMD~\cite{pmd} and compilers to detect and correct errors in code snippets~\cite{reidnlp2tc, terragni2016csnippex, 10.1145/3520304.3528772, 10.1145/3520304.3533970}. The process of converting code into another lower-level language is more complicated than just generating an AST (compilers parse code as only one step of the compilation process), meaning that they report errors that parsing alone does not. Because Java code must be compiled before it can be run, the ability to compile a snippet is a useful measure of quality in a reuse context -- code that does not compile is thus not runnable. Many compilers, such as the Eclipse Java compiler and the \ts\ compiler, are designed to report multiple errors and are used to report error information within an IDE. To the best of our knowledge, no existing work attempts to use the \ts\ compiler to evaluate and fix errors in Node.js code. Based on these observations, we devise NCC to fill this gap.

\subsection{Automatic Code Correction}

Much work on fixing code has focused on software bugs in runnable programs, evaluated via test cases. In contrast to this, code correction in the context of code reuse deals with fragmented, often unrunnable code, where these approaches cannot be applied. To solve the problem of correcting unrunnable snippets, existing tools rely on static analysis to identify errors and inform heuristic fixes. CSnippEx~\cite{terragni2016csnippex}, for Java, employs an existing suite of fixes from Eclipse, while NCQ~\cite{reidNCQ} does the same in Node.js using ESLint's fixes. NLP2TestableCode~\cite{reidnlp2tc} in Java uses a set of custom heuristic fixes, and is able to insert missing import statements and variable definitions. Jigsaw~\cite{cottrell2008Jigsaw}, another Java tool, allows developers to supply a method to integrate and a destination class or function, then extracts structural information to make integration changes. Where it cannot automatically fix integration errors, it inserts comments and highlights parts of code for developer attention. 

Besides NCQ, other work in \js\ and Node.js looks at repairing software bugs in runnable code. Vejovis~\cite{ocariza2014vejovis} automatically suggests repairs for DOM-based \js\ faults to developers, but these repairs require runnable code and are not applicable for Node.js. The use of AI models to fix code is also of interest: Lajkó et al.~\cite{lajko2022towards} look at the use of the GPT-2 model to fix software bugs; after training the model to fix \js\ bugs, they found that it did so correctly in most cases. AI tools that generate code snippets, such as GitHub Copilot~\cite{copilot}, a plug-in for VSCode that uses OpenAI's more advanced GPT-3-based Codex~\cite{codex}, are able to generate snippets that match the surrounding context, eliminating the need to integrate snippets. However, there is some concern about the quality of the output of these systems, with regard to bugs, vulnerabilities, and correctness for given queries~\cite{dakhel2022github, nguyen2022empirical}. NCC aims to build on existing work on correcting errors in Node.js code snippets, by combining \ts{}'s existing fix suite, with custom heuristic fixes, as well as utilising comments where developer intervention is still needed. Additionally, we hypothesise that better error detection using a compiler will enable more accurate error correction.


\subsection{Code Deletion}
Code deletion, for example, at the granularity of lines or statements, is a unary operator that is easy to implement. It also does not require any code analysis or synthesis and can be a component of a more complex operation, such as replacing a line with another. Therefore, code deletion is typically included in studies related to code improvement~\cite{legoues2012operators,harrand2019neutralvariants,10.1145/3520304.3533970,10.1145/3520304.3528772, petke2019survey, ginelli2022deletions}. Often, these studies report errors that are \textit{fixed} (relative to a given test suite) by removing the offending code. 

For the problem of correcting unrunnable code where test cases cannot be used, line deletion, in combination with error reporting, has previously been used as a solution. For example, NLP2TestableCode~\cite{reidnlp2tc}, which is an Eclipse plug-in that assists in reuse of Stack Overflow code, employs a line deletion algorithm as the last step in a suite of fixes, using the Eclipse compiler to provide error information. The NCQ code reuse tool~\cite{reidNCQ} implements a similar functionality for Node.js, using ESLint to evaluate errors. 
Similarly, Licorish and Wagner~\cite{10.1145/3520304.3528772} combined static analysis with the Gin genetic improvement framework~\cite{10.1145/3321707.3321841} (which includes deletion operations, among others) to improve Java code snippets on Stack Overflow. In contrast to these works, we investigated the possibility of using the \ts\ compiler to inform line deletions as just one of the potential ways it could be used for code corrections.

\section{Approach}

\begin{figure}[h]
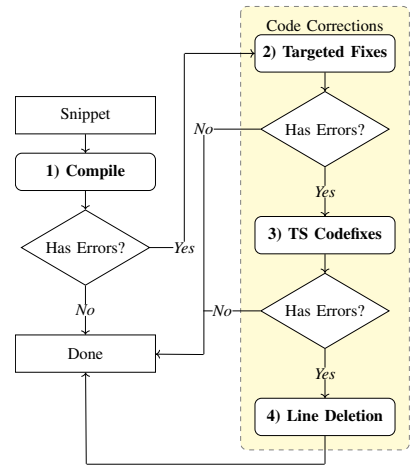

\centering
\includestandalone[width=0.6\linewidth]{figures/overview}
\caption{The NCC pipeline.}
\label{fig:pipeline}
\end{figure}

Node Code Correction (NCC) has four stages, illustrated in \autoref{fig:pipeline}; 1) compilation to identify errors; 2) targeted fixes; 3) \ts\ codefixes and finally 4) line deletion. Each snippet is initially compiled to check for errors; then, for erroneous snippets, the code correction process begins. First, a series of heuristic custom fixes are attempted; if errors continue to exist, then \ts{}'s builtin codefixes are applied where available. Finally, line deletion is employed to handle remaining errors. After any change, the code is compiled again to update the error information. This section describes each aspect of NCC.

\subsection{Identifying Errors}

Identifying errors is the first step to correcting errors. To do this, NCC uses the \ts\ compiler from version 4.9.4
of the \ts\ package. To optimise compilation speed, the \ts\ compiler is run programmatically and in-memory for a given string of code, using a custom \texttt{CompilerHost} that handles the interface between the compiler and the `file system'. This custom \texttt{CompilerHost} stores the code string as an in-memory \texttt{sourceFile}, instead of looking on the file system. Furthermore, we cache any required files (such as \ts\ definition files) between compiles. By default, the \ts\ compiler will check all loaded files for errors, considerably slowing down the compilation time, so we specify only looking at our single input file. The \ts\ compiler is then isolated from other code, by only allowing it to access files in the `typescript' and `@types/node' folders that are needed for \ts\ to function.

We configured the compiler with options to match a default v18.16.0 Node.js environment for example, we enabled Node.js types and did not allow JSX (JavaScript XML) because the dataset is meant to be Node.js code only. To trigger the \ts\ compiler's \js\ mode, the in-memory file was named with a `.js' file ending. Due to the size of the dataset, for this investigation of the entire NPM registry, we do not try to install each snippet's source package, though the \ts\ compiler is capable of deriving additional type information that could have enabled more accurate type information. Additionally, the need to install packages for each snippet would increase the time to fix, which is one of the benefits of static analysis like this. We simply ignore the `Cannot find module' error on \texttt{require} statements, and the compiler will then continue to generate general Node.js errors. On compilation, the compiler generates a list of diagnostics, including error code, message, start location, and length. To deal with rare cases where the compiler threw an error or never finished compiling, we run the compiler in a separate process with a timeout of 60 seconds.

\subsection{Targeted Fixes}

On the basis of our experimentation with prototype versions of NCC, we devise a series of custom heuristic fixes that address common errors that other stages cannot correct. Our heuristics for identifying and correcting errors thus embody a series of iterative enhancements that integrate lessons learned from these early prototypes. 

We created two fixes for the common error \texttt{Cannot find name}, which occurs for undeclared variables. We identify that the cause of the error is either a missing `require' for a package, or a variable not being defined. Our two fixes thus are to insert the import or define a placeholder variable. We leverage the \ts\ compiler's ability to generate ASTs for even erroneous code to provide information about the context and surrounding code, and to keep track of errors that exist on the same line as each other. We ignore cases of the \texttt{Cannot find name} error where it may be reported for non-code (for example, terminal commands): in cases of \texttt{expression expected} and \texttt{unexpected keyword or identifier} we presume that the code on that line has additional issues, and we make no changes. 

For missing require statements, we check if the identifier could be an API usage, and then check if the name matches to a built-in library. For these cases, we insert a \texttt{require} statement. Undefined functions are ignored to allow \ts{}'s codefixes to handle these cases instead. In other, non-function, cases, we attempt to get the expected type of the undeclared variable. That is: if the identifier is an argument of a function, we check for the expected type from the parent function. From here, we can insert a placeholder string, number or array of strings or numbers. Where a type cannot be determined, we default to a string. Additionally, for more complex types, we default to a placeholder string with a comment noting the suggested type. These placeholders serve to move the snippet to a more `correct' state, while indicating where developer intervention may be needed. The motivating example in \autoref{sec:example} demonstrates a case in which both of these fixes are applied. After applying a fix, we compile to see if the changes do not increase the total number of errors; if not, the change is kept.

\subsection{\ts\ Codefixes}

We employ \ts{}'s codefixes (the Quick Fix suggestions that \ts\ can provide to an IDE, for example, when integrated with VSCode) to automatically correct errors. \ts\ codefixes require use of the \texttt{LanguageService} API, not the compiler, but, similarly to the \ts\ compiler, we speed up runs via in-memory objects and caching. Sharing a \texttt{DocumentRegistry} object between runs, and only updating the input `file' for each snippet, gives considerable speed benefits. \ts\ supports fixes for 1,190 of its 1,878 available error types.

We adapt the codefix procedure of the Microsoft \texttt{ts-fix} tool~\cite{tsfix}, which automatically fixes errors in \ts\ projects. For each error, a set of \texttt{CodeFixAction}s is supplied if they exist, each with its own set of changes that must be made to the text. All possible changes are combined into a list, sorted by the earliest start and then the smallest change. Then, we filter the list to remove changes that would overlap (i.e., affect the same part of the string), before applying them to the text. Then we compile the code again to update the error count. 

\subsection{Line Deletion}

\DontPrintSemicolon
\begin{algorithm}
    $S_{best} \gets \text{Initial snippet}$\;
    \tcp{Step 1: Get errors}
    $S_{best}\text{.}errors \gets  Compile(S_{best})$\;
    $\text{done} \gets false$\;
    $\text{$errorNo$} \gets 0$\;
    \While{$done == false$}{
        $S_{current} \gets S_{best}$\;
        \tcp{Step 2a: Check if done}
        \uIf{$errorNo >= S_{current}.errors.length$}{
            $done \gets true$\;
        }
        \tcp{Step 2b: Try delete error}
        \Else{
            $S_{current}\text{.}DeleteLineFor(errorNo)$\;
            $S_{current}\text{.}errors \gets  Compile(S_{current})$\;
            \tcp{Step 4a: Keep deletion }
            \uIf{$S_{current}.errors <= S_{best}.Errors$}{
                $S_{best} \gets S_{current}$\;
                $errorNo \gets 0$\;
            }
            \tcp{Step 4b: Try next error}
            \Else{
                $errorNo++$\;
            }
        }
    }
    \KwRet{$S_{best}$}\;
\caption{Line Deletion Algorithm}
\label{alg:deletion}
\end{algorithm}

Line deletion is a commonly used technique to reduce errors in code snippets~\cite{legoues2012operators,harrand2019neutralvariants,10.1145/3520304.3533970,10.1145/3520304.3528772, petke2019survey, ginelli2022deletions, reidNCQ}. We adapt the NCQ line deletion algorithm to work with the \ts\ compiler. The deletion algorithm functions as illustrated in \autoref{alg:deletion}, and is run on snippets that still have errors after the codefix stage. The algorithm attempts to find the `best' snippet based on error count, by deleting lines affected by errors. The `deletion' occurs by commenting out the line, just like in NCQ; in a code reuse situation these erroneous lines may still be useful to a developer by providing additional context, and commented out code can aid developers in fixing bugs, debugging and adding features~\cite{pham2020secret}. 
We prefer line deletion over statement deletion for the issue of code fragments, as not all snippets are parsable.

First, the snippet $S$ is compiled to find errors. If there are no errors, the process stops there. If there are errors, the algorithm starts with the first error and attempts to delete the associated line. The new snippet is then re-evaluated, and if the error count did not increase, the deletion is kept and the $errorNo$ variable is reset as the error list is now changed. If the change made things worse, we revert the change and move on to the next error. The loop ends when there are no errors or all errors have been processed, and the algorithm returns the snippet with the least errors it can produce. In some cases, snippets are commented out completely (so-called `empty' snippets) in order to reduce the snippet to zero errors.

The major change from NCQ is that \ts\ reports more than a single `failed here' parsing error, unlike ESLint. This means that NCC's line deletion algorithm is capable of trying multiple changes when one does not work. Furthermore, because the mined dataset still contains some non-Node.js snippets even after filtering, we handle additional edge cases not shown in the algorithm based on the unexpected behaviour of the \ts\ compiler. We ignore the previously discussed crashing snippets. Additionally, it is possible for an error to persist even on a commented-out line, so we check if the line has been commented out and skip it. In cases where the reported error location exceeds the actual snippet length, which we interpret as a problem parsing the snippet, we terminate the line deletion process.

\section{Dataset}
\label{sec:dataset}

This section provides an overview of the dataset used to evaluate the performance of the code correction tool. The dataset consists of two main sources: NPM snippets and Stack Overflow edits, each described in detail in their respective subsections. `Snippet' in both instances refers to code fragments mined directly from either markdown or HTML, by looking for code blocks. We do not attempt any combination of related snippets in a single source, as developers and tools often treat snippets as self-contained, even when that may not be true.

\subsection{NPM Snippets}
\label{sec:npm_snippets}

Reid et al.~\cite{reidNCQ} originally ran NCQ's code corrections over a dataset of 2,161,911 code snippets mined from the NPM registry as of May 2021. The dataset contains snippets extracted from markdown code blocks in the package READMEs. Heuristics were employed to ensure the dataset was filtered for only Node.js snippets, manually verified on a sample of 384 READMEs (confidence level 95\%, confidence interval 5). However, non-JavaScript code snippets (including terminal commands, \ts\ and JSX, a JavaScript extension used for React) may still be present within the dataset.

For our evaluation, we use the same publicly available dataset for our evaluation. However, because NCQ is a Node.js REPL, it implements some REPL-specific rules and fixes to make reusing code snippets in this environment easier. Because of this, we rerun its correction on the dataset after disabling these rules to better emulate the scenario described in \autoref{sec:example} (a developer looking to reuse code snippets in a regular Node.js programming environment) and report errors before and after fixes.


\subsection{Stack Overflow Edits}
\label{sec:so_edits}

In order to compare NCC's performance to how developers manually edit code, we evaluate on a set of Stack Overflow snippets, for which we have the first and most recent edit. We used the December 2020 version of the SOTorrent dataset~\cite{DBLP:conf/msr/BaltesDT008, baltes2020annotated}, retrieving the code-only \texttt{PostBlockVersion}s for all accepted answers of posts tagged `Node.js', giving us a total of 299,389 snippet versions, or `edits', across 182,205 snippets. For our SOEdits dataset, we look only at snippets where there are at least two versions, there was some change between the first and last version, and the first version has at least one error, creating a dataset of 21,431 snippet `before' and `after' edit pairs. These pairs represent an original erroneous snippet, and the current, edited snippet on SO.

\begin{table}[h]
    \centering
    \caption{Summary of SOEdits dataset.}
    \begin{tabular}{l|r}
        \toprule
        All snippets & 182,205 \\
        All versions & 299,389 \\
        \midrule
        &\\
        All SOEdit pairs & 21,431 \\
        Improvement only & 15,969\\
        Fixed only & 1,099\\
        \bottomrule
    \end{tabular}
    \label{tab:SOEdits_dataset}
\end{table}

By running the \ts\ compiler, we observe that this set of snippet pairs does not necessarily represent an improvement over time; overall, the number of errors increased between edit pairs, as did the number of lines of code. For 74.51\% of SO edit pairs (15,969 pairs), there was an improvement in errors between edits, and for only 5.12\% of edits (1,099 pairs), all errors were corrected. For this reason, we further filter the dataset to the 15,969 snippets that show improvement and create an additional subset for the 1,099 snippets that were `fixed'. \autoref{tab:SOEdits_dataset} shows the breakdown of the data.

\section{Evaluation}
\label{sec:eval}

We run both NCQ and NCC's corrections on a dataset of 2,161,911 code snippets from NPM package documentation (described in \autoref{sec:npm_snippets}) and record results at each stage. To establish baseline data of what errors ESLint and \ts\ can identify, we also run only the error reporting. Experiments were run with the latest LTS version of Node.js as of April 2023 (18.16.0), version 4.9.4 of \ts\ and version 8.31.0 of ESLint. We make the assumption that developers looking for code online expect it to be up-to-date and compatible with the recommended version of Node.js. We configure the error reporters in NCQ and NCC (ESLint and \ts{}) comparatively to emulate the scenario described in \autoref{sec:example}; a code snippet pasted into an empty file, in an otherwise empty Node.js project, with no packages installed. Additionally, our error reporters are configured for CommonJS, or `script' mode, where \texttt{require} statements are used to import packages and top-level await is not allowed. Because NCQ's corrections were designed for its REPL context, REPL-specific rules and fixes were disabled so as not to impact results. Although the dataset may still contain non-Node.js code snippets despite filtering, such as \ts\ and JSX, we limit the evaluation to Node.js; where TS and ESLint have options to process this code without errors, we do not enable them. We ask the following research questions:

\begin{adjustwidth}{0.5em}{0.5em}
\vspace{1ex}
\noindent
\textbf{RQ1.} \rqone

\noindent
\textbf{RQ2.} \rqtwo

\noindent
\textbf{RQ3.} \rqthree

\noindent
\textbf{RQ4.} \rqfour
\vspace{1ex}
\end{adjustwidth}
\noindent

Furthermore, to evaluate the NCC results against the way developers manually fix errors, we compare the results with the set of improvements in the SOEdits dataset, described in \autoref{sec:so_edits}. We ask the following research question:

\begin{adjustwidth}{0.5em}{0.5em}
\vspace{1ex}
\noindent
\textbf{RQ5.} \rqfive
\vspace{1ex}
\end{adjustwidth}
\noindent

\subsection{\rqone}
\label{sec:rq1}

We ask this question to characterise the frequency and types of errors in NPM package documentation and also to establish a baseline to compare our corrections. We ran the \ts\ compiler on all 2,161,911 code snippets and found that only 569,201 code snippets (26.3\%) had no errors. \ts\ identified a total of 14,707,149 errors in the set, an average of 6.8 errors per snippet. Looking at only erroneous snippets, the average number increases to 9.2.

\begin{figure}[t]
    \centering
    \includegraphics[width=0.8\linewidth]{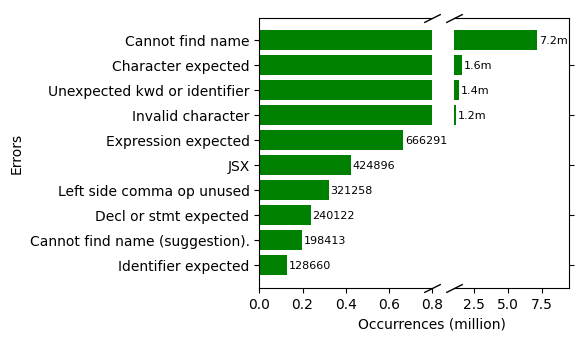}
    \caption{Most common error types in NPM documentation, reported by TS.}
    \label{fig:common_err}
\end{figure}

\ts\ reports 404 different error types on our dataset. Almost half of the 14.7 million errors \ts\ detects are for the error type \texttt{cannot find name}, with 7.2 million occurrences; this is visible in \autoref{fig:common_err}. This error reports cases where an identifier was referenced without a declaration. There is a similar, but separately numbered, error that suggests an alternative name, where a misspelling is suspected, accounting for another 198,413 errors. The second most common error type, \texttt{character expected}, accounts for 1.6 million errors. With the exception of the \texttt{JSX} error which appeared in 154,737 code snippets (7.16\%), common errors can be characterised as missing or unexpected characters, keywords, identifiers, statements, or expressions. Error messages typically provide the error-causing token or the expected token. \ts\ is also able to detect when functions do not exist on a type; the \texttt{Property does not exist on type} error is the 12th most common with 83,483 occurrences.

\begin{figure}[h]
    \begin{lstlisting}
var prompt = require('prompt');

prompt.start();
prompt.get(['username', 'email'], function (err, result) {
    console.log('Command-line input received:')
    console.log('  username: ' + result.username)
    console.log('  email: ' + result.email)});
    \end{lstlisting}
    \begin{lstlisting}
const {username, email} = await prompt.get(['username', 'email']);
    \end{lstlisting}
    \caption{Two code snippets from the package \texttt{prompt}.
    }
    \label{fig:prompt}
\end{figure}

\autoref{fig:prompt} shows a common situation in NPM package documentation. These two code snippets from the README for the \texttt{prompt} package demonstrate two ways to get input from a user on the command-line using a prompt; using a callback or using \texttt{await}. However, the second code snippet would generate a \texttt{cannot find name} error when evaluated by the \ts\ compiler. The variable \texttt{prompt} is undeclared in the second snippet, but not in the first.

Code snippets are often not intended to be working examples but rather to demonstrate functionality; they often omit code that would be repeated between code snippets, such as the \texttt{require} statement in \autoref{fig:prompt}. However, developers and automated tools still use them this way. The number of \texttt{cannot find name} errors in the dataset suggests that missing variables are common and that this practice is widespread.

\vspace{0.5cm}
\fbox{\begin{minipage}{0.9\linewidth}
\textbf{Summary:}~The majority (73.7\%) of code snippets in NPM package documentation have some kind of error. On average, the snippets have 6.8 errors. The most common error was for undeclared variables.
\end{minipage}}
\vspace{1em}

\subsection{\rqtwo}
\label{sec:rq2}

To compare the two error reporters, we ran ESLint on the same snippets. We use a modified version of the configuration from NCQ, with the REPL specific errors disabled. We disable `linting' rules concerning formatting and only look at errors that would affect code functionality.

ESlint reports a similar rate of erroneous snippets, with only 26.3\% of snippets having no errors. However, we observe that the average number of errors per snippet sits at 1.26 errors (and 1.71 errors for the erroneous set). In fact, ESLint reports only a fraction (18.5\%) of the errors that \ts\ can on the same set. This is because of the 2,722,241 errors reported, 27.8\% are parsing errors. 

\begin{figure}[h]
    \centering
    \includegraphics[width=0.8\linewidth]{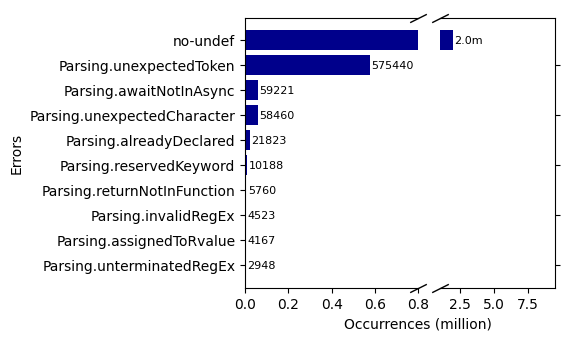}
    \caption{The 10 most common error types via ESLint.}
    \label{fig:eslint_common_err}
\end{figure}

ESLint reports 185 different error types, of which 175 of the types (94.6\%) are parsing errors. \autoref{fig:eslint_common_err} reflects this, where all but the most common error reported by ESLint (no-undef at 2m occurrences) are parsing errors, such as unexpected tokens (575,440 occurrences) and use of \texttt{await} outside of an async function (only allowed in ES modules), at 59,221 instances. The prevalence of parsing errors is an issue for two reasons. First, ESLint reports only a single parsing error per snippet indicating why parsing failed, thus these snippets do not generate an AST, nor can ESLint run its rule detection or fixes. This means for 47.46\% of snippets with errors, we are only able to detect a single, unfixable error. Secondly, ESLint rule detection besides from `no-undef' accounts for only 2,211 errors. \ts{}'s `unexpected token' error, for example, occurs for only 60,885 occurrences; instead, \ts\ has an increase in other more specific error types that might provide more useful information on the cause of the error. The types of errors ESLint can report are limited due it's intended purpose as a \emph{linter}; further lowing the error rate, the majority of rules are not enabled by NCQ as they are not useful for identifying erroneous code. In this context Thus, the error information is incomplete, and ESLint may not necessarily be useful for evaluating runnability or informing fixes.

\vspace{0.5cm}
\fbox{\begin{minipage}{0.9\linewidth}
\textbf{Summary:}~ ESLint reports considerably fewer errors than the \ts\ compiler -- an average of 1.3 vs.~6.8 per snippet. 47.46\% of erroneous snippets have a single error where parsing failed, resulting in an average of 1.71 errors per erroneous snippet compared to \ts's 9.2. In these cases, it also cannot generate an AST to enable fixes. These results indicate that ESLint is limited in what it can tell us about code.
\end{minipage}}
\vspace{1em}

\subsection{\rqthree}
\label{sec:rq3}

First, we look at the impact of only the \ts\ codefixes on the set of snippets. After applying the codefixes, the number of snippets without errors increased from 569,201 (26.3\%) to 648,814 (30.0\%). The total number of errors was reduced from 14,707,149 to 14,096,112 (a decrease of 4.2\%). In total, 602,629 snippets (27.9\%) had changes made to fix errors.

\begin{figure}[h]
    \centering
    \includegraphics[width=0.8\linewidth]{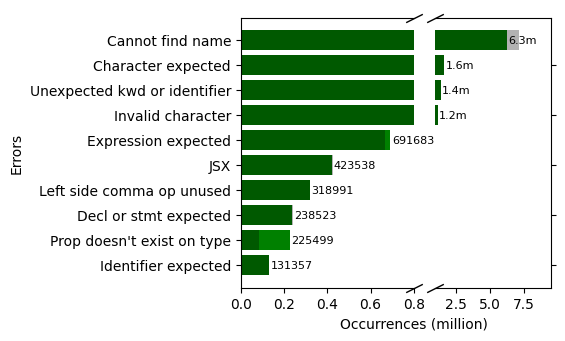}
    \caption{The 10 most common error types after TS codefixes. Light Grey represents a decrease in errors from previous results in \autoref{fig:common_err}, where light green represents an increase.}
    \label{fig:common_err_ts}
\end{figure}

\autoref{fig:common_err_ts} illustrates the most common error types after codefixes and shows significant changes. All 10 of the most common errors in \autoref{fig:common_err} had fixes, however, not all fixes can be applied to every error, and some fixes can introduce new errors. The most common error is still \texttt{Cannot find name}, but it has reduced from 7.2 million to 6.3 million occurrences. The error type \texttt{Cannot find name (suggestion)} no longer appears in the most common errors, reducing from 198,313 occurrences to just 25,979. As discussed in \autoref{sec:rq1}, this error type directly relates to its suggested fix, so the reduction is logical here. However, the error type \texttt{Property doesn't exist on type} increased by 142,016 occurrences. \texttt{Expression expected} also visibly increased. The other errors have minor increases/decreases that are not visible on this scale. The increase in error \texttt{Property doesn't exist on type} probably results from fixes for undefined variables, where a variable is defined that does not make sense on the basis of usage.




Next, we consider the impact of line deletion in combination with TS codefixes. We find that the number of snippets without errors increased from the original 569,201 (26.3\%) to 1,622,272 (75.0\%), an increase of 185.0\%. The total number of errors that could not be fixed also decreased from the original 14,707,149 to 925,277. The average number of errors per snippet decreased to 0.43. In total, 1,343,992 snippets (62.2\% of the total) had changes made to reduce errors. 

The line deletion stage accounts for a 150.0\% improvement, or additional 973,458 error-free snippets, from the \ts\ codefix stage. However, it also comments out all lines for 483,169 code snippets (22.3\% of the total set and 49.6\% of snippets it makes error-free). To measure the impact of the line deletion algorithm, we counted the number of lines before and after deletion. In total, 4,031,366 lines of code were commented out, 22.1\% of the total lines of code. Combined with the lines added from \ts\ codefixes, there are 2,902,083 fewer lines after NCC's corrections. However, deleted lines are only commented out, so they can still be useful to developers by providing additional context and guiding them with what to do next.

\begin{figure}[h]
    \centering
    \includegraphics[width=0.8\linewidth]{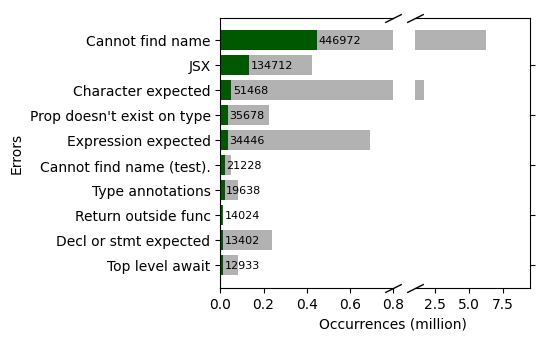}
    \caption{The 10 most common error types after deletion and TS codefixes. Shading represents improvement over \autoref{fig:common_err_ts}.}
    \label{fig:common_err_all}
\end{figure}

The types of errors that were reported changed considerably between codefixes and deletion, as seen in \autoref{fig:common_err_all}. The original 7.2 million instances of \texttt{cannot find name} were again reduced to 446,972 instances. New errors now populate the top 10: \texttt{cannot find name (it)} where a testing library using the function \texttt{it()} was not installed, \texttt{Type annotations} for the use of \ts, and \texttt{Top level await}. All common errors saw reductions from their previous values, except \texttt{Return outside function}, which increased by 7,728 occurrences after line deletion.

\begin{figure}[h]
\begin{lstlisting}[language=diff]
+ var s = "YOUR VALUE HERE";
var words = s.split(" ");
    \end{lstlisting}
    \caption{Example of undeclared variable with no fix, and a proposed fix.}
    \label{fig:cannotfind}
\end{figure}

Based on these results, we implement a limited suite of targeted fixes for the most common error that still persists: \texttt{cannot find name}. We conjecture that the ability to define variables will enable NCC to reduce the number of line deletions, and thus empty snippets. Despite the error type \texttt{Cannot find name} having TS codefixes in some cases, seemingly `simple' cases such as the example in \autoref{fig:cannotfind} cannot be fixed. On the basis of these cases, we introduce custom fixes.

\begin{figure}[h]
    \centering
    \includegraphics[width=0.8\linewidth]{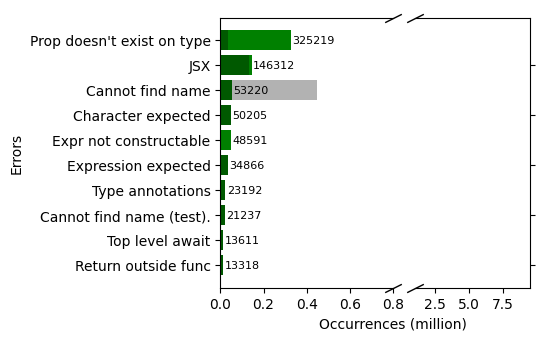}
    \caption{The 10 most common error types after all fixes. Shading represents prior results from \autoref{fig:common_err_all}.}
    \label{fig:common_err_cf}
\end{figure}

We run all fixes on the dataset and observe that the number of snippets with no errors decreases slightly by 1,929 (a decrease from 75.0\% to 74.94\%). However, this does not tell the entire story: the total number of errors fell by 2.76\% to 899,774, and the number of empty snippets fell considerably from 22.35\% of the dataset to 7.41\%. \autoref{fig:common_err_cf} shows how some errors increase but that there is a considerable decrease in \texttt{Cannot find name}. We see an increase in the error \texttt{property doesn't exist on type}, likely due to the addition of placeholder definitions which default to strings in many cases. Similarly, the increasing error \texttt{expression not constructable} also handles a similar case, where the previously undefined identifier should instead be a constructable object.

\vspace{0.5cm}
\fbox{\begin{minipage}{0.9\linewidth}
\textbf{Summary:}~NCC improves the number of error-free snippets by 184.67\%, and most of the remaining erroneous snippets have some changes to reduce errors. However, 7.41\% of the snippets are entirely commented out by the corrections. The results indicate that leveraging \ts\ enables NCC's custom fixes to decrease errors, but that additional heuristic fixes could further reduce the reliance on `last resort' line deletion.
\end{minipage}}
\vspace{1em}

\subsection{\rqfour}
\label{sec:rq4}

We ask this question to investigate whether the use of a compiler like \ts\ instead of a linter can improve error-informed code corrections. NCQ's code correction approach consists of three components: evaluating errors using ESLint, fixing errors using ESLint's built-in fixes, and the line deletion algorithm which runs on parsing errors. We record errors before and after all correction steps.

With all fixes, we find that NCQ's corrections are able to increase the number of snippets without errors from 569,419 to 982,832 (45.46\% of total code snippets), an increase of 413,413 code snippets (72.6\%). Furthermore, the total number of reported errors (that could not be fixed) actually increases by 34,502, as the correction of parsing errors enable more ESLint rule violations to be detected. In line with this, the average errors per erroneous snippet increased to 2.34. 113,621 of the 413,413 snippets made error-free (27.48\%) had all lines commented out, and the line deletion algorithm removed 3,461,047 lines (20.2\% of all lines).

\begin{figure}[h]
    \centering
    \includegraphics[width=0.8\linewidth]{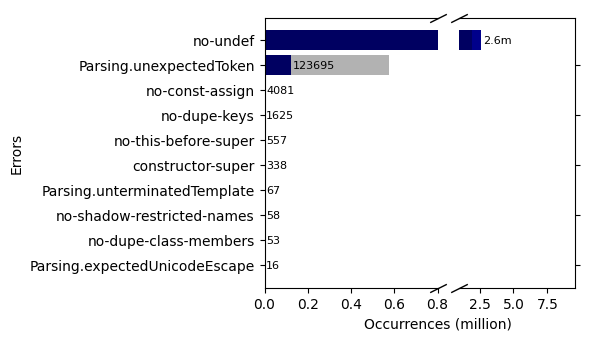}
    \caption{The 10 most common error types via ESLint after NCQ's corrections. Shading represents the initial values before fixes (see \autoref{fig:eslint_common_err}).}
    \label{fig:eslint_common_err_fixes}
\end{figure}

\autoref{fig:eslint_common_err_fixes} shows the most common errors after fixes. Errors reduced for parsing errors as expected, with \texttt{unexpected tokens} now at 123,695 occurrences. Now that so many parsing errors have been corrected with the deletion algorithm, other errors appear in the top 10. We see that \texttt{no-undef} saw an increase to 2,626,174 occurrences, as NCQ employs no fixes for this error, and other non-parsing errors are now visible such as \texttt{no-const-assign}. We note from our results that ESLint's built-in fixes have very little impact on the dataset: after the ESLint fix stage, errors only reduced by 2 and a single snippet was made error-free. ESLint's fixes mostly solve formatting issues and only work on parsable snippets, so this is expected. We compare these results with NCC's use of \ts\ codefixes, which corrected all errors in 79,613 snippets and resulted in a 611,037 error reduction.

It is difficult to compare the impact of each approach on the quality of code snippets; we do not attempt to run snippets after fixes to see if there is an improvement in runnability, and neither evaluator can give us a `correct' number of errors. We also cannot compare the exact number of errors between approaches, as each evaluator reports errors differently. However, we can see that NCQ results in more empty snippets as part of its corrections, and that it does not correct as many snippets. Because most erroneous snippets with ESLint have a single parsing error, this means that if a line deletion does not improve the code snippet, there are no alternatives to try. Empty snippets and number of deleted lines also reduced between approaches: 14.29\% of corrected snippets were empty for NCC (7.41\% of the dataset), compared to NCQ's 27.48\% (14.33\%). This suggests that the use of \ts\ enables more accurate line deletion and that additional fixes reduce the reliance on deletion.

\vspace{0.5cm}
\fbox{\begin{minipage}{0.9\linewidth}
\textbf{Summary:}~Compared to NCQ's code corrections, we find that NCC has a higher improvement rate and that NCC leaves fewer snippets empty (7.41\% vs 14.33\%). We find that ESLint's automated fixes implemented in NCQ had little effect on improving code snippets, only fixing 1 snippet.
\end{minipage}}
\vspace{1em}

\subsection{\rqfive}
\label{sec:rq5}

To compare NCC to manual fixes, we evaluate against Stack Overflow snippet pairs, for which we have an original snippet containing at least one error (the `pre-edit' snippet), and the most recent with reduced errors (the `post-edit' snippet). We look at both the total `improvement' set, and the 1,099 subset where all errors were fixed. Though we do not expect our limited suite of fixes to correct all 1,099 snippets, the aim of the comparison is to see how well NCC performs despite this.

\begin{figure}[h]
    \centering
    \includegraphics[width=0.8\linewidth]{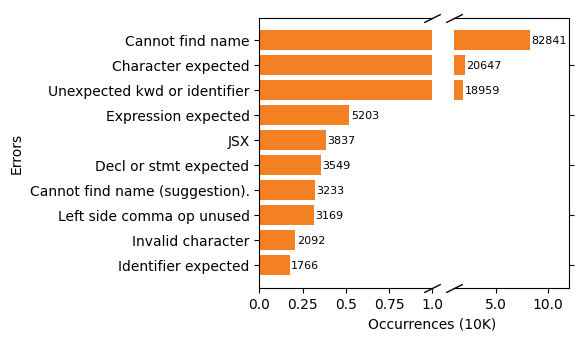}
    \caption{The 10 most common errors for Stack Overflow Edits pre-edit.}
    \label{fig:so_errors}
\end{figure}

First, we observe the error landscape of the pre-edit `improvement' set. The \ts\ compiler reports a total of 160,602 errors in the dataset: an average of 10.06 errors per snippet. This set of snippets only contains erroneous snippets, so all snippets have at least one error. \autoref{fig:so_errors} shows the most common error types before editing, with the most common error remaining `cannot find name' like in the previous results. Again, Stack Overflow snippets often miss elements that might exist in other snippets or in the question.

\begin{figure}[h]
    \centering
    \includegraphics[width=0.8\linewidth]{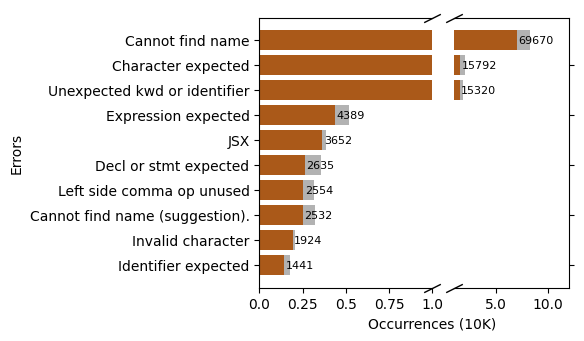}
    \caption{The 10 most common error types for Stack Overflow Edits post-edit. Grey represents the change between edits.}
    \label{fig:so_errors_post}
\end{figure}

Next, we observe the reduction in errors after manual fixes, represented in the `post-edit' set. We see a 17.60\% reduction in total errors and 6.88\% of the snippets have all errors corrected.  \autoref{fig:so_errors_post} shows the change in error types, which can be summarised as a general reduction in all common error types.

\begin{figure}[h]
    \centering
    \includegraphics[width=0.8\linewidth]{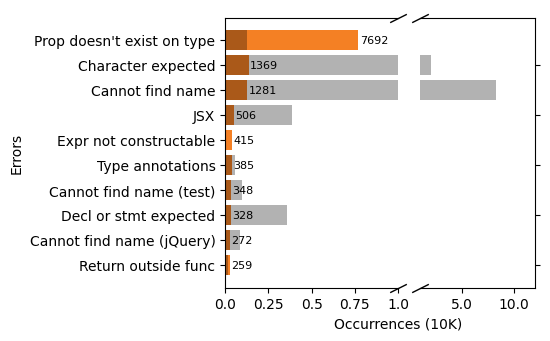}
    \caption{The most common errors for Stack Overflow snippets after NCC.}
    \label{fig:so_errors_ncc}
\end{figure}

When we run NCC over the original snippets, we observe that our suite of fixes enables a 90.69\% decrease in errors. 7,469 snippets are made error-free: an additional 6,370 snippets over the post-manual-edit set, representing 46.77\% of the dataset. This result is achieved with a 5.71\% rate of empty (`commented out') snippets, compared to only 3 (0.27\%) for manual edits. \autoref{fig:so_errors_ncc} shows how the error landscape changes: The occurrences of \texttt{cannot find name}, previously the most common error, reduce to only 1,281 occurrences. Again, we see similar errors increase as in \autoref{sec:rq3}. However, we check for an increase in errors after making changes to ensure that the change does not make the code worse, and the value is intended to be modified by developers with their own value.

For the subset of 1,099 snippets that were manually fixed in their post-edit version, NCC is able to fix 726, 66.06\% of the set. 47 (4.28\%) of the snippets are entirely commented out by NCC's fixes.

\vspace{0.5cm}
\fbox{\begin{minipage}{0.9\linewidth}
\textbf{Summary:}~
NCC can resolve all errors for 46.77\% of SO snippets, reducing errors by 90.69\% with a rate of 5.71\% empty snippets. Evaluated against snippets with manual corrections, NCC can fix 66.06\% of these snippets, which is a promising result. 
\end{minipage}}
\vspace{1em}

\section{Threats and Limitations}

There are several potential threats to the validity of this study. Firstly, our results on quality and correctness of the code snippets are based on reported errors from both \ts\ and ESLint, but neither of these tools can accurately represent the \emph{runnability} of the code. Furthermore, we did not try to install each package within our dataset, but because \ts\ can gather type information and use it to report errors, this could have provided additional error information. We assume that the inability to parse or compile a snippet relates to its quality and runnability, which is true for compilable languages like Java but may not necessarily hold for Node.js. Similarly, because we do not run code snippets, either before or after fixes, we cannot know the impact of fixes on runnability. Our fixes for Stack Overflow snippets may report fewer errors, but our automatic fixes may not be similar to the kinds of fixes that developers produce manually. Additionally, removing lines may reduce errors at the cost of expected behaviour. Because we focus on lines, and not statements, errors over multiple lines may not easily be addressed by our algorithm.

Snippets were mined devoid of context, in order to replicate developer copy-paste and code recommender systems like NCQ, but this method may account for some of the missing variable errors. Additionally, while care was taken to limit the mined datasets to only Node.js code, as described in \autoref{sec:dataset}, non-code still exists in the dataset and may impact results. Finally, the results of our evaluation are specific to Node.js and the version used, and we cannot claim that they generalise to other languages, or even versions of Node.js. Additionally, there are limitations to our approach. NCC simply re-implements \texttt{ts-fix}'s batch approach to applying \ts{}'s fixes, which does not validate each fix individually. Like the line deletion algorithm and heuristic fixes, each change could be checked via the compiler to ensure that no fix makes a snippet worse. Furthermore, heuristic fixes can always be further refined to handle more situations. We acknowledge the limitations of such fixes, in that they must be individually designed for each error case and make guesses about missing parts of code. We look with interest at Large Language Models (LLMs) like OpenAI's Codex and ChatGPT that might provide new AI solutions for this problem. GitHub's Copilot plug-in already generates snippets in the editor for a given task and code context and could change code reuse practices. However, there are concerns about the quality of generated code. Future work may investigate how a similar system can be applied to existing code snippets.

\section{Conclusion and Future Work}
\label{sec:conclusion}

Developers often rely on code snippets found online for reference and assistance in their projects. However, most of these snippets are not runnable, requiring developers to spend additional time fixing errors, which can be especially challenging when using third-party libraries. Existing approaches to automatically identify and fix errors in snippets have primarily focused on static analysis using parsers and linters, as snippets often lack test cases or do not run. Although these techniques have proven useful in some cases, there is still a need for a better way to evaluate and correct errors in Node.js code. Our work aims to address this gap by using the \ts\ compiler for more effective and accurate code correction compared to a linter like ESLint, which on average reports only one error per snippet. 

Our results indicate that the \ts\ compiler enables more effective and accurate code identification and correction when compared to ESLint. The \ts\ compiler is also capable of detecting more errors and more informative errors, and its built-in fixes affect more snippets. Additionally, the reported error information and ASTs generated by the \ts\ compiler enable the use of additional heuristic fixes on more snippets. Based on these results, we suggest the use of the \ts\ compiler for static analysis on Node.js datasets over linters, and the NCC approach for automating code reuse.

Future work could integrate NCC within a code recommendation system or as a IDE plug-in, to find, insert and then correct code snippets from online, similar to NCQ. Additionally, we could study how useful developers find the fixes generated by NCC, either by asking them questions about the changes or having them use the tool in a code reuse scenario. It may also be interesting to investigate how accurately \ts\ errors correlate with runnability, when trying to run code snippets, and how well NCC makes the snippets runnable.

\section*{Acknowledgement}

Brittany’s research was supported by an Australian Government Research Training Program (RTP) Scholarship. 

\IEEEtriggeratref{16}
\bibliographystyle{IEEEtran}
\bibliography{main.bib}

\end{document}

%% file: packages.tex
\usepackage{cite}
\usepackage{amsmath,amssymb,amsfonts}
\usepackage{algorithmic}
\usepackage{graphicx}
\usepackage{textcomp}
\usepackage{xcolor}
\usepackage{changepage}
\usepackage{soul}
\usepackage{enumitem}
\usepackage[utf8]{inputenc}
\usepackage{booktabs}
\usepackage[english]{babel}
\usepackage[ruled,noend]{algorithm2e}
\usepackage{hyperref}
\usepackage{color}
\usepackage{listings}
\usepackage{tikz}
\usepackage{standalone}
\usepackage{xcolor}

%% file: macros.tex
\newcommand{\ts}{TypeScript}
\newcommand{\js}{JavaScript}

\newcommand{\rqone}{What errors does \ts\ detect in NPM documentation?}
\newcommand{\rqtwo}{How does error detection differ between ESLint and \ts?}
\newcommand{\rqthree}{What is the impact of NCC on the set of NPM snippets?}
\newcommand{\rqfour}{How does NCC compare to NCQ's code corrections?}
\newcommand{\rqfive}{How does NCC compare to manual fixes?}

\hypersetup{
    colorlinks=true,
    linkcolor=blue,
    anchorcolor=blue,
    filecolor=blue,      
    citecolor=blue,
    urlcolor=blue
}

\definecolor{lightgreen}{rgb}{0.8,1,0.8}
\definecolor{lightyellow}{rgb}{1,1,0.8}
\definecolor{lightred}{rgb}{1,0.8,0.8}
\definecolor{baseColour}{RGB}{255, 128, 85}
\definecolor{ncqColour}{RGB}{118, 165, 175}
\definecolor{ncqColour2}{RGB}{162, 196, 201}

\SetCommentSty{mycommfont}
\SetAlFnt{\small}
\SetAlCapFnt{\small}
\SetAlCapNameFnt{\small}
\algsetup{linenosize=\tiny}

\lstdefinelanguage{javascript}{
    keywords={typeof, new, true, false, catch, function, return, null, catch, switch, var, if, in, while, do, else, case, break},
    keywordstyle=\bfseries,
    ndkeywords={class, export, boolean, throw, implements, import, this},
    ndkeywordstyle=\color{darkgray}\bfseries,
    identifierstyle=\color{black},
    sensitive=false,
    comment=[l]{//},
    morecomment=[s]{/*}{*/},
    commentstyle=\color{blue}\ttfamily,
    stringstyle=\color{purple}\ttfamily,
    morestring=[b]',
    morestring=[b]"
}
\newcommand{\lstbg}[3][0pt]{{\fboxsep#1\colorbox{#2}{\strut #3}}}
\lstdefinelanguage{diff}{
    morecomment=[f][\lstbg{lightgreen}]+,
    morecomment=[f][\lstbg{lightred}]-,
    morecomment=[f][\lstbg{lightyellow}]\\,
    keywords={typeof, new, true, false, catch, function, return, null, catch, switch, var, if, in, while, do, else, case, break},
    keywordstyle=\bfseries,
    ndkeywords={class, export, boolean, throw, implements, import, this},
    ndkeywordstyle=\color{darkgray}\bfseries,
    identifierstyle=\color{black},
    sensitive=false,
    morecomment=[s]{/*}{*/},
    commentstyle=\color{blue}\ttfamily,
    stringstyle=\color{purple}\ttfamily,
    morestring=[b]',
    morestring=[b]"
}
\addto\extrasenglish{%
}

\usetikzlibrary{
    arrows.meta,
    chains,
    scopes,
    positioning,
    shapes.geometric
}
\pgfdeclarelayer{background}
\pgfdeclarelayer{foreground}
\pgfsetlayers{background,main,foreground}

\tikzset{
    processDiagram/.style={
        startstop/.style={
            rectangle,
            draw,
            minimum width=3cm, 
            minimum height=0.8cm,
            join = by arrow,
            fill = black!0,
        },
        decision/.style = {
            diamond,
            aspect=1.7,
            draw,
            minimum width=2.5cm, 
            minimum height=1cm, 
            align=center,
            join = by arrow,
            fill = black!0,
        },
        process/.style = {
            rectangle, 
            rounded corners, 
            draw,
            minimum width=3cm,
            minimum height=0.8cm, 
            align=center,
            join = by arrow,
            fill = black!0,
        },
        process2/.style = {
            rectangle, 
            rounded corners, 
            draw,
            minimum width=3cm,
            minimum height=0.8cm, 
            align=center,
            join = by arrow,
            fill = ncqColour2,
        },
        line/.style = {
            -
        },
        invisible/.style = {
            join = by line,
            minimum width=0cm, 
            minimum height=0cm,
            inner sep=0pt,
        },
        invisititle/.style ={
            font = \bfseries,
            join = by arrow,
            outer sep = 5pt,
        },
        title/.style = {
            font = \bfseries,
            fill = black!20,
            join = by arrow,
            minimum width = 3cm,
            minimum height=0.8cm,
            draw,
        },
        output/.style ={
            join = by line,
            minimum width=0cm, 
            minimum height=0cm,
            inner sep=0pt,
            font = \itshape,
            align = center,
        },
        arrow/.style = {
            ->
        },
    }
}

%% file: figures/overview.tex
\begin{tikzpicture}[processDiagram, node distance=4mm and 5cm]

{ [start chain=trunk going below]
    \node [startstop, on chain] (n0) {Snippet};
    \node [process, on chain] (n1) {\textbf{1) Compile}};
    \node [decision, on chain] (errors) {Has Errors?};
    {[start branch = c going below]
        \node [output, on chain = going right, xshift=-4.5cm] (c0o) {Yes};
        \node [invisible, on chain = going above, yshift=3.63cm] (i0) {};
        \node [process, on chain = going right, xshift=-3.5cm] (c1) {\textbf{2) Targeted Fixes}};
        \node [decision, on chain] (errors2) {Has Errors?};
        {[start branch = cc going left]
            \node[output, on chain, xshift=4cm] {No};
            \node[invisible, on chain = going below, yshift=-4.3cm] (i1) {};
        }
        \node [output, on chain] (c1o) {Yes};
        \node [process, on chain = going below] (c2) {\textbf{3) TS Codefixes}};
        \node [decision, on chain] (errors3) {Has Errors?};
        {[start branch = cc going left]
            \node[output, on chain, xshift=4.4cm] {No};
            \node[invisible, on chain = going left, xshift=4.8cm] {};
        }
        \node [output, on chain] (c2o) {Yes};
        \node [process, on chain] (c3) {\textbf{4) Line Deletion}};
        \node [invisible, on chain, yshift=-0.2cm] (i2) {};
        \node [invisible, on chain = going left, xshift=-0.1cm] (i3) {};
    }
    \node [output, on chain] (n1o) {No};
    \node [startstop, on chain] (n2) {Done};
    \draw [arrow] (i1) -> (n2);
    \draw [arrow] (i3) -> (n2);
}

\begin{pgfonlayer}{background}
    \path (c1.west |- c1.north)+(-0.3,0.6) node (a) {};
    \path (c3.east |- c3.south)+(+0.3,-0.3) node (b) {};
    \path[fill=yellow!20,rounded corners, draw=black!50, dashed](a) rectangle (b); 

    \node[above of = c1, yshift=0.2cm]{Code Corrections};
\end{pgfonlayer}

\end{tikzpicture}